\documentclass[pdflatex,sn-mathphys-num]{sn-jnl}
\usepackage{amsmath,amssymb,amsfonts,bm}
\usepackage{graphicx}
\usepackage{booktabs}
\graphicspath{{figures_qip/}}
\newcommand{\openone}{\mathbb{1}}

\newcommand{\ket}[1]{|#1\rangle}
\newcommand{\bra}[1]{\langle#1|}
\newcommand{\e}{\mathrm{e}}
\newcommand{\ii}{\mathrm{i}}
\newcommand{\sx}{\sigma_x}\newcommand{\sy}{\sigma_y}\newcommand{\sz}{\sigma_z}
\newcommand{\tx}{\tau_x}\newcommand{\tz}{\tau_z}

\begin{document}

\title[Helper-assisted quantum sensing with low dressed-state leakage]{Helper-assisted quantum sensing with low dressed-state leakage}

\author*[1]{\fnm{Shirwan} \sur{Abdullah}}\email{shirwan.abdullah@kissr.edu.krd}

\affil*[1]{\orgname{Kurdistan Institution for Strategic Studies and Scientific Research}, \orgaddress{\city{Sulaymaniyah}, \postcode{46001}, \state{Kurdistan Region}, \country{Iraq}}}

\abstract{We study a two-qubit sensing protocol in which only a helper qubit is read out. Calibrated loading and unloading encode a transverse signal in the helper's phase, keep leakage out of a selected dressed-state band small, and return the sensor close to its idle ground state. We derive an information bound for a single sensor whose excitation stays below a prescribed budget, and simulate repeated operation with an uncertain coupling and finite-duration resets in a driven open-system model. At the tested settings, finite measurement records support estimation with infrequent resets. A helper flip at zero coupling, combined with a reversal of the coupling sign, suppresses a static helper offset shared across a record while retaining the signal, at a cost of about 27\% of the no-offset information of a plain sequence of equal duration. Under slowly varying noise, averaging each cycle's likelihood over the correct noise law removes most of the added bias, and a wrong noise law can worsen the estimate. A selected directly read sensor with ideal controls and repeated nondemolition verification has the higher information rate. Any practical benefit of avoiding direct readout therefore requires a device-specific disturbance model.}


\maketitle

\section{Introduction}

Quantum sensors encode an unknown signal in a probe state and infer it from a measurement, with realisations ranging from spins in diamond and trapped ions to superconducting circuits and semiconductor quantum dots \cite{Degen2017}. The quantum Fisher information sets a local precision limit through the Cram\'er--Rao inequality for unbiased estimators \cite{Braunstein1994,Paris2009}, and noise limits the information that can be retained \cite{Huelga1997,Escher2011,Demkowicz2012}. The readout step is often the most invasive part of the cycle. Direct readout can disturb the sensor, depending on the measurement basis and on the device, by heating it, changing its charge or acting on a nearby register; for charge-sensor readout of spin qubits, for instance, the sensor's back-action has been modelled to leave a mixed post-measurement state and to cause leakage out of the computational subspace \cite{Svastits2026}, and the leakage induced by dispersive readout of a superconducting qubit varies by more than an order of magnitude between readouts of equal assignment fidelity \cite{Hazra2025}. Reading a second system coupled to the sensor may avoid that disturbance.

Several established approaches motivate this work. Strongly interacting or engineered dressed states protect a sensing subspace \cite{Dooley2018,Gorecki2026}, and protective measurement by a qubit probe has been analysed for a chain of measurements with a Lindblad environment, with the protecting Hamiltonian kept on throughout \cite{Schlosshauer2020}; neither treats the repeated loading and unloading of the dressing, nor the leakage it leaves behind, nor the estimation from a finite record. Ancilla-assisted estimation, in which a meter coupled to the probe by a Hamiltonian term is read locally, has been analysed for thermometry \cite{Kiilerich2018}, for the preparation of a probe ensemble by a weakly coupled qubit \cite{Chen2026} and with repeated measurement and reset of the ancilla while the probe accumulates the signal \cite{Mentesoglu2026}; none of these places a budget on the excitation left in the probe. Ancilla-assisted frequency up-conversion enables transverse dc magnetometry with low-frequency noise rejection \cite{Liu2019}, and ancilla-based readout in tweezer clocks has demonstrated repeated phase detection and conditional sensor reset \cite{TweezerClock2024}. Repeated quantum nondemolition (QND) readout of the sensor itself, with adaptive verification \cite{Braginsky1980,Vaartjes2026}, is the direct-sensing reference against which any indirect scheme must be costed. Our first paper on this subject \cite{Abdullah2026} studied a stationary charge detector biased below the gap of an interacting quantum-dot sensor to limit the sensor's excitation during readout. Here we study a pulsed protocol that stores the signal in a coherent helper phase; its protection criterion distinguishes leakage out of the dressed band during sensing from the excitation left in the sensor after unloading, and the sensor need not stay close to its idle ground state during the pulse.

The model is a sensor qubit with a fixed splitting $\Delta$ and a transverse signal $\theta$, coupled to a helper qubit through a tunable $\sx\tz$ interaction. We combine calibrated loading and unloading with conditional resets for repeated operation, under an explicit leakage convention. We derive an information bound for a sensor working alone under an excitation budget, and compare the helper protocol with it at equal duration and leakage threshold while keeping the different reference projectors of the two explicit. The simulations include the stated control, readout and reset times, and the estimation uses finite records and averages over the uncertain coupling. A helper flip at zero coupling, paired with a reversal of the coupling sign, suppresses a static helper offset shared across a record while preserving the signal, and we test it against slowly varying offsets as well. Finally we compare the information per unit of laboratory time with a selected directly read sensor that verifies its readout by repeated QND reads.

The helper does not have the higher rate in that comparison. Under its ideal analysis controls, the direct reference collects about 10\% more information at the baseline duration and 25--42\% more at the duration of the echo. The helper's possible benefit is the avoidance of direct sensor readout, and the present results quantify the cost of that choice; establishing a practical advantage requires a model of the disturbance caused by that readout, which we do not supply.
\section{Model and protocol}
\label{sec:model}

\begin{figure}
\centering\includegraphics[width=\textwidth]{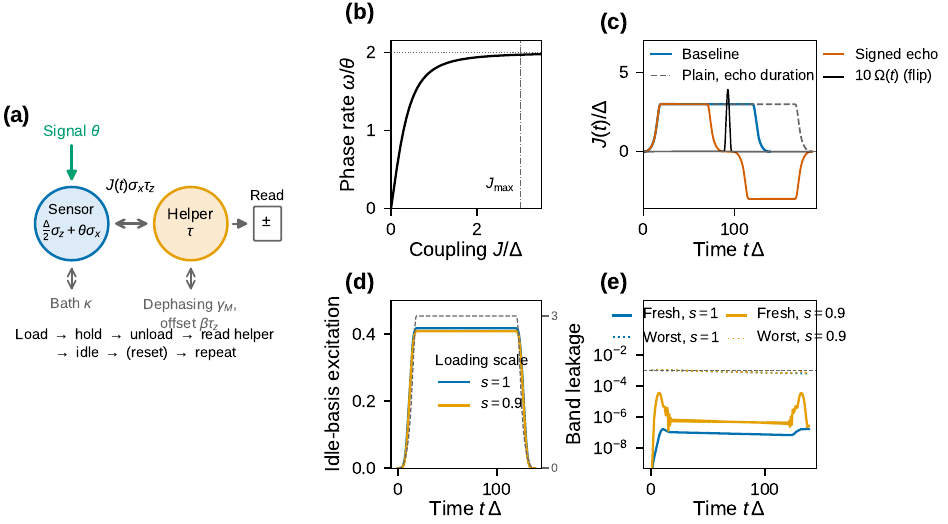}
\caption{The protocol and its protection. (a) A sensor qubit with splitting $\Delta$ and a transverse signal $\theta$ is coupled to a helper qubit through a tunable $J(t)\sx\tz$ interaction; only the helper is read. The sensor sees a bath of strength $\kappa$; the helper suffers Markovian dephasing $\gamma_M$ and, in Sec.~\ref{sec:echo}, a slow offset $\beta\tz$. (b) Rate $\omega$ at which the relative phase of the two dressed branches grows, divided by the signal, as a function of the coupling; it saturates at $2$ for $J\gg\Delta$. (c) Coupling waveforms: the baseline pulse (ramps of $19.3$ time units, hold $100$), the plain pulse of the echo's duration (hold $146.5$), and the signed-echo sequence of Sec.~\ref{sec:echo}, in which a helper flip $\Omega(t)\tx$ of eight time units is applied at $J=0$ between two holds of opposite coupling sign. (d) Excitation of the sensor measured in its idle basis along one baseline pulse in the open-system model, at $\theta=0.010$ for the nominal loading scale $s=1$ and for $s=0.9$; the dashed line is $J(t)$. (e) Leakage out of the reference band along the same pulse, for a fresh input in the sensor ground state (solid) and for the worst admissible input carrying excitation $10^{-3}$ (dotted); the dashed line marks $10^{-3}$. Time is in units of $1/\Delta$ throughout.}
\label{fig:scheme}
\end{figure}

We consider a sensor qubit with a fixed splitting $\Delta$ and a weak transverse signal $\theta$, coupled to a helper qubit $\tau$ through a controllable interaction that is longitudinal for the helper and transverse for the sensor [Fig.~\ref{fig:scheme}(a)]. With $\hbar=\Delta=1$ the Hamiltonian is
\begin{equation}
\hat H(t)=\tfrac{1}{2}\sz+\theta\,\sx+J(t)\,\sx\tz+\hat H_{\rm cd}(t),
\label{eq:H}
\end{equation}
where $\sigma$ and $\tau$ are Pauli operators of the sensor and the helper, $J(t)$ is the coupling waveform, and $\hat H_{\rm cd}$ is a counterdiabatic correction defined below. The signal $\theta$ is the quantity to be estimated; its prior is uniform on $[0.006,\,0.014]$, centred on $\theta_c=0.010$, a signal of one per cent of the splitting known to $\pm40\%$. Table~\ref{tab:settings} collects the settings and separates the inference prior from the narrower set of signals the controller considers and from the values at which the truth was simulated.

The interaction commutes with $\tz$, so the two helper states $\tz=\pm1$ see different sensor Hamiltonians, $\hat H_\pm=\tfrac12\sz+(\theta\pm J)\sx$, which we call the two branches. Their ground energies are $E_{g,\pm}=-\tfrac12\sqrt{1+4(\theta\pm J)^2}$, and the relative phase between the branches grows at the rate
\begin{align}
\omega(\theta,J)&=E_{g,-}-E_{g,+}\nonumber\\
&=\tfrac12\Big[\sqrt{1+4(\theta+J)^2}-\sqrt{1+4(\theta-J)^2}\Big],
\label{eq:omega}
\end{align}
which is odd in $\theta$ and, for $\theta\ll J$, equals $2b(J)\theta$ with $b(J)=2J/\sqrt{1+4J^2}$ [Fig.~\ref{fig:scheme}(b)]. At the coupling we use, $J_{\max}=3$, the gain is $2b=1.97$ and $\omega=0.0197$ at $\theta_c$. The idea is thus a Ramsey experiment on the helper: the helper is prepared in $\ket{+}$, the sensor is loaded adiabatically into the ground state of its branch, the helper accumulates the phase $\int\omega\,dt$ during the ramps and the hold at $J_{\max}$, the sensor is unloaded, and the helper is read along a fixed equatorial axis, at the angle that maximises the single-pulse Fisher information at the centre of the prior ($2.313$ rad from the $x$ axis for the baseline pulse; Appendix~\ref{app:cycle}). The sensor is never measured. Note that during the hold the sensor is in a dressed ground state, not its idle one; in the idle basis its excitation is $0.42$ at $J_{\max}=3$ [Fig.~\ref{fig:scheme}(d)]. The protocol protects a band, and the quantities we keep small are the leakage out of that band and the excitation left behind after unloading.

The band is defined by the reference projector $\hat P_{\rm band}(t)=\sum_{s=\pm}\ket{g_s(t)}\bra{g_s(t)}\otimes\ket{s}\bra{s}$ onto the instantaneous ground states of the two branch Hamiltonians at the true coupling, without the counterdiabatic terms. Loading and unloading are made slow, over ramps of $\tau_r=19.27$ time units each, and are corrected by a transitionless-driving term \cite{Berry2009,GueryOdelin2019}, $\hat H_{\rm cd}=c_0(t)\sy+c_Z(t)\sy\tz$ with $c_0\pm c_Z=\tfrac12\dot\varphi_\pm$ and $\tan\varphi_\pm=2(\theta_c\pm J)$ the branch mixing angles at the centre of the prior; the hold lasts $100$ and the whole pulse $T=138.54$ [Fig.~\ref{fig:scheme}(c), Appendix~\ref{app:pulse}]. Because the correction is designed at $\theta_c$ and at the nominal coupling, it is imperfect for other signals and for a miscalibrated coupling, and the residual leakage is one of the quantities we quantify. (A counterdiabatic term that is itself robust to parameter uncertainty has been proposed for sensing during state preparation \cite{Morawetz2026}; here the correction serves only to make the loading and unloading transitionless.)

Two kinds of imperfection are kept throughout. The coupling is known only up to a scale factor $s$, $J\to sJ$, with $s\in[0.9,1.1]$; the scale enters the branch phase, so it is an uncertain parameter that the inference must average over. The helper is noisy: its coherence decays at the Markovian rate $\gamma_M=10^{-3}$ (jump operator $\sqrt{\gamma_M/2}\,\tz$), and in Sec.~\ref{sec:echo} it also carries a slow offset $\beta\tz$, which a plain Ramsey sequence cannot distinguish from the signal. The sensor couples to a bosonic bath with strength $\kappa=10^{-3}$ (Sec.~\ref{sec:open}); the helper readout is binary, along a preset axis, with an assignment error of one per cent.

Repeated operation is organised in cycles. A controller forecasts, from its own model of the sensor state evolved over a small set of plausible signals and coupling scales (Table~\ref{tab:settings}), whether the coming pulse would drive the excitation of the worst candidate above a threshold of $10^{-3}$, and if so applies an engineered damping reset before loading; after the pulse and the helper readout, a second check triggers a recovery reset when the post-readout estimate exceeds the threshold. Damping is its only corrective action. Resets cost time and are counted in all rates we quote; every operation and its assigned duration are listed in Table~\ref{tab:ops}.

\begin{table}
\caption{Settings of the model, the controller and the inference. The three signal sets are distinct: the prior is what the estimator assumes, the controller candidates are the hypotheses it protects against, and the truth values are the points at which the simulations were run. Energies are in units of $\Delta$ and times in units of $1/\Delta$.}
\label{tab:settings}
\footnotesize
\centering\begin{tabular}{p{0.35\textwidth}p{0.54\textwidth}}\toprule
signal prior (inference) & uniform on $[0.006,\,0.014]$, grid step $10^{-4}$ \\
controller signal candidates & $\{0.008,0.009,0.010,0.011,0.012\}$ \\
controller scale candidates & $\{0.9,\,1.0,\,1.1\}$ \\
simulated truths, baseline & 17 pairs, $\theta\in[0.008,0.012]$, $s\in[0.9,1.1]$ \\
simulated truths, echo & $(\theta,s)=(0.010,1.0)$, $(0.012,0.9)$ and $(0.008,1.1)$ \\
simulated truths, direct reference & $\theta=0.008$, $0.010$, $0.012$ (no scale); helper rows at $s=1$ \\
loading-scale prior & uniform on $[0.9,1.1]$, 9 nodes (3 for echo runs) \\
coupling, gain & $J_{\max}=3$, $2b=1.97$ \\
counterdiabatic design point & $\theta_c=0.010$, $s=1$ \\
helper dephasing & $\gamma_M=10^{-3}$ \\
sensor bath & Ohmic, $\kappa=10^{-3}$, $\omega_c=20$, $k_BT_b=0.1$ \\
helper readout & binary, preset axis, assignment error $0.01$ \\
excitation threshold & $10^{-3}$ (forecast and post-readout) \\
\bottomrule
\end{tabular}
\end{table}
\section{A bound for a sensor working alone}
\label{sec:bound}

What could a sensor do by itself under a comparable restriction, namely that its excitation above a reference ground state stays small at all times? A nondegenerate ground state carries no phase reference, so a sensor that is never excited learns about $\theta$ only through the static susceptibility of its ground state, which does not grow with time. Excitation buys a phase reference, and the question is how much information a given excitation budget buys. The answer is a simple bound, stated here and proved in Appendix~\ref{app:bound}.

Consider a single qubit with fixed splitting $\Delta$ under $\hat H(t)=\hat H_{\rm ref}(t)+u_y(t)\sy$ with $\hat H_{\rm ref}=\tfrac12\Delta\sz+[\theta+u_x(t)]\sx$, where $u_x$ and $u_y$ are arbitrary bounded functions of time chosen without knowledge of $\theta$. The sensor is prepared in the ground state of $\hat H_{\rm ref}(0)$, evolved coherently for a time $T$ and measured in any way. Let $\epsilon_t$ be the population of the excited state of $\hat H_{\rm ref}(t)$ at time $t$, and suppose $\epsilon_t\le\epsilon$ throughout. Then the quantum Fisher information for $\theta$ obeys
\begin{equation}
F_Q(T)\;\le\;4\left(\frac{3}{\Delta}+A+2\sqrt{\epsilon}\,T\right)^{2},
\label{eq:bound}
\end{equation}
where $A=\int_0^T\|\dot{\hat B}(t)\|\,dt$ is the total variation of $\hat B=\Delta\sy/[\Delta^2+4(\theta+u_x)^2]$, i.e.\ of $\Delta/\Omega^2$ with $\Omega$ the instantaneous gap of $\hat H_{\rm ref}$; $A\le2/\Delta$ when the gap rises and falls once, i.e.\ for a dressing $u_x$ that grows and then returns without changing the sign of $\theta+u_x$; a dressing that passes through zero, or a control that keeps re-dressing the sensor, makes $A$ and the bound grow. The $y$ control does not enter $A$ because it commutes with $\hat B$, but it does enter $\epsilon_t$, which is measured against $\hat H_{\rm ref}$ and not against the full field. With a signal-independent preparation the constant $3/\Delta$ becomes $2/\Delta$.

The structure of the bound is the point. At large $T$ and bounded $A$ the information can grow at most as $16\epsilon T^2$, quadratically in time but with a prefactor set by the permitted excitation; a cross term $\propto\sqrt{\epsilon}\,T$ and a constant remain at finite time. The leading coefficient is approached, not attained. For a sensor prepared at the design signal in the dressed basis of a strong field $u_x$, kicked suddenly to excitation $\epsilon$ and held there, the leading long-time term of the information is $16\epsilon(1-\epsilon)b^2T^2$ with $b=2(\theta_c+u_x)/\sqrt{\Delta^2+4(\theta_c+u_x)^2}<1$ (the single-sensor analogue of the helper gain $b(J)$); at $T=58.5$ and $\epsilon=10^{-3}$ the exact numerical value is $52.7$, against $53.3$ for that leading term and $54.8$ for $16\epsilon T^2$. This construction has a signal-independent preparation, the $2/\Delta$ case of the bound. Smooth six-harmonic controls remain far below the bound: the best values found by a penalised local search at $\epsilon=10^{-4}$, $10^{-3}$ and $2\times10^{-3}$ are $4.7$, $7.3$ and $18.0$ against bounds of $72$, $186$ and $321$ at $T=58.5$ (Appendix~\ref{app:bound}); they are the best values found, not optima, and they lie far below the bound because keeping $\epsilon_t$ small at every instant is a severe restriction on a single qubit.

\begin{figure}
\centering\includegraphics[width=0.62\textwidth]{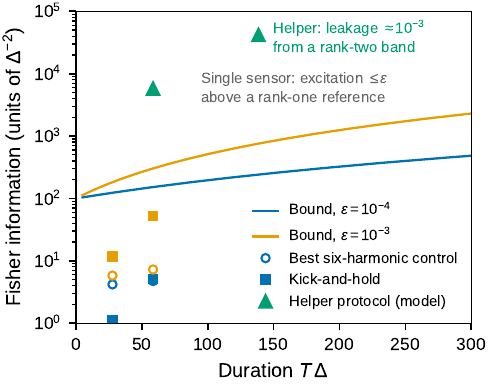}
\caption{Information against duration under an excitation budget. Lines: the bound of Eq.~(\ref{eq:bound}) at $A=2/\Delta$ for budgets $\epsilon=10^{-4}$ (blue) and $10^{-3}$ (orange), for a single sensor whose excitation is measured against its rank-one reference ground state. Filled squares: the kick-and-hold construction; open circles: the best six-harmonic controls found by a penalised search at the same budgets. Triangles: the helper protocol in the open-system model at holds 20 and 100, at the nominal signal and coupling, whose constraint is instead a maximum leakage of approximately $10^{-3}$ out of a rank-two band at the nominal settings. The two constraints use the same numerical threshold relative to different projectors.}
\label{fig:bound}
\end{figure}

The single-sensor theorem restricts excitation relative to a rank-one reference ground state. The helper protocol instead uses a rank-two dressed band that retains a relative phase between its branches: the signal enters with opposite sign in the two branches, while longitudinal noise on the sensor enters both nearly equally. We compare the two at the same numerical leakage threshold and duration, keeping these different constraints explicit. For $\epsilon$ we take the helper's own band leakage, maximised over the pulse and over all admissible inputs (sensor excitation up to $10^{-3}$), which is approximately $10^{-3}$ at the nominal signal and coupling in the open-system model of Sec.~\ref{sec:open} (and up to $1.2\times10^{-3}$ over the tested coupling range, Sec.~\ref{sec:open}). At the nominal signal and coupling the helper collects $5\,781$ units of information per pulse at a hold of 20 ($T=58.5$) and $41\,950$ at a hold of 100 ($T=138.5$), against single-sensor bounds of $303$ and $758$: factors of 19 and 55 (Fig.~\ref{fig:bound}). These are nominal-point, constrained comparisons between a coherent theorem and an open-system model prediction. They do not imply superiority over direct protocols that leave the reference ground state during sensing, including the nondemolition reference of Sec.~\ref{sec:direct}, and they are not an unrestricted metrological advantage.
\section{Open-system model}
\label{sec:open}

The sensor is coupled through $\sz$ to a bosonic bath with an Ohmic spectral density, an exponential cutoff at $\omega_c=20\Delta$, temperature $k_BT_b=0.1\Delta$ and coupling strength $\kappa=10^{-3}$, so the idle sensor is cold. Because the control fields vary on time scales comparable to the bath correlation time, a static Lindblad equation is not adequate. We use the universal Lindblad equation of Nathan and Rudner \cite{Nathan2020,Nathan2021erratum,Nathan2024}, in which the jump operator is built from the bath correlation function and the driven propagator over a memory window, so that the dissipator follows the drive; the accompanying Lamb shift is included. The equation is completely positive by construction, and its accuracy parameter \cite{Nathan2020} is $\Gamma\tau_B=0.016$ here. Its accuracy has been benchmarked against the exactly solvable damped oscillator beyond ultraweak coupling \cite{TelloBreuer2024} and bounded rigorously for many-body systems under bulk dissipation \cite{Ikeuchi2025}; neither setting covers the driven use made here, with interventions between pulses. The helper's coherence decays at the Markovian rate $\gamma_M$ through the jump operator $\sqrt{\gamma_M/2}\,\tz$. Grids, memory windows and the construction of the sequence channels are given in Appendix~\ref{app:numerics}.

Four checks fix the status of this model. The small leakage and end-point values are converged in the numerics: halving the propagation step changes the end-point excitation and the maximum band leakage at the nominal point by $0.07\%$ and the information per pulse by $5\times10^{-7}$ relative, and extending the memory window by two thirds changes none of them at four digits (Table~\ref{tab:conv}, Appendix~\ref{app:numerics}). The information per pulse agrees with a time-convolutionless second-order treatment of the same bath to within $0.011\%$ at the three loading scales tested at hold 100 \cite{Breuer2002}; building the dissipator for an isolated pulse or for one embedded in a periodic train changes the channel elements by at most $3\times10^{-6}$; and including the static Lamb shift of the idle sensor in the idle and reset intervals changes the information per cycle at the nominal point by less than one part in $10^5$ (it is included in all production results). The first check is a statement about the discretisation; the other three support the internal consistency of a weak-coupling driven model. None establishes a microscopic description of every intervention.

Three approximations limit the model. The endpoint excitations are predictions of one completely positive approximation, without the drive-dependent corrections to the dissipator that arise beyond second order in the system--bath coupling \cite{Bernazzani2025}. Measurement and reset interventions are applied as separately generated maps between pulses rather than inside the memory integrals. For the echo sequences the dissipator is built with the helper offset $\beta$ set to zero, which changes the sequence channel by at most $8\times10^{-10}$ per element at the two tested offsets $\beta=\pm0.02$ (an observed difference, not a bound).

Figure~\ref{fig:scheme}(d,e) shows what the model predicts along one pulse. The sensor's excitation in its idle basis rises to $0.42$ during the hold and returns to $1.7\times10^{-7}$ at the end ($2.7\times10^{-7}$ at $s=0.9$): the protocol protects the dressed band and the end point, not the bare ground state throughout. The leakage out of the reference band for a fresh input never exceeds $1.7\times10^{-7}$ at the nominal signal and coupling, $2.6\times10^{-7}$ at the nominal coupling and the tested signals $0.008$ and $0.012$, and $4.4\times10^{-5}$ at any tested pair of those signals with $s\in\{0.9,1,1.1\}$ (the maximum occurs at the shorter hold of 20); at $s=0.9$ the counterdiabatic correction no longer cancels the nonadiabatic transitions in the ramps, which is the origin of the $3.5\times10^{-5}$ visible in the figure. The net change of the excitation of a fresh input over a pulse is below $2\times10^{-5}$ in magnitude, and for the worst admissible input, which already carries excitation $10^{-3}$, the leakage stays at $1.0$--$1.2\times10^{-3}$: the pulse adds little to an excitation it inherits.
\section{Repeated operation and estimation}
\label{sec:cycle}

\begin{table}
\caption{Operations of one cycle, the duration assigned to each and how it is counted, in the helper protocol and in the directly read reference of Sec.~\ref{sec:direct}. Times are in units of $1/\Delta$. The reset is a damping map toward the sensor ground state at the true signal, a model assumption; the controller and the estimator propagate each of their own hypotheses with the reset map at that hypothesis, and the true signal is not supplied to them. Information is the expected Fisher information of each readout given the conditional state, with the derivative taken at the true signal and fixed loading scale, summed over every readout of a history; rates are ratios of pooled totals over 400 histories (information summed over histories divided by elapsed time summed over histories), with bootstrap intervals over histories.}
\label{tab:ops}
\footnotesize
\centering\begin{tabular}{p{0.18\textwidth}p{0.36\textwidth}p{0.36\textwidth}}\toprule
operation & helper protocol & directly read reference (QND verification) \\
\midrule
initial sensor state & maximally mixed, then three resets ($3\times0.4605$) & same \\
decision interval & $0.05$, bath only; forecast reset if predicted excitation $>10^{-3}$ & $0.05$, bath only; no forecast \\
sensing pulse & ramps $19.27$ each, hold $100$ ($T=138.54$); helper in $\ket{+}$ & preparation $5$, ramp, hold, unload ($T=138.54$), $J_d=1.5$ \\
analysis rotation & none (helper read along a preset axis) & instantaneous ideal unitary, fixed at $\theta_c$ \\
readout & helper, binary, error $0.01$; zero assigned duration & sensor, $m$ reads of duration $t_{\rm read}=1$ each, error $0.01$ \\
 & helper re-prepared in $\ket{+}$ at zero cost & block accepted when all $m$ agree, else repeated (cap 12 blocks) \\
return & none & instantaneous ideal swap of the energy basis if the accepted report is ``excited'' \\
idle interval & $3$, bath only & $3$, bath only \\
recovery reset & $0.4605$ (rate 10, removes 99\%) if post-readout estimate $>10^{-3}$ & none for $m\ge2$ \\
abort / halt & abort cycle if forecast fails twice; halt history if recovery fails twice, or if an aborted cycle leaves the worst candidate above threshold & not applicable \\
\bottomrule
\end{tabular}
\end{table}

\begin{figure}
\centering\includegraphics[width=\textwidth]{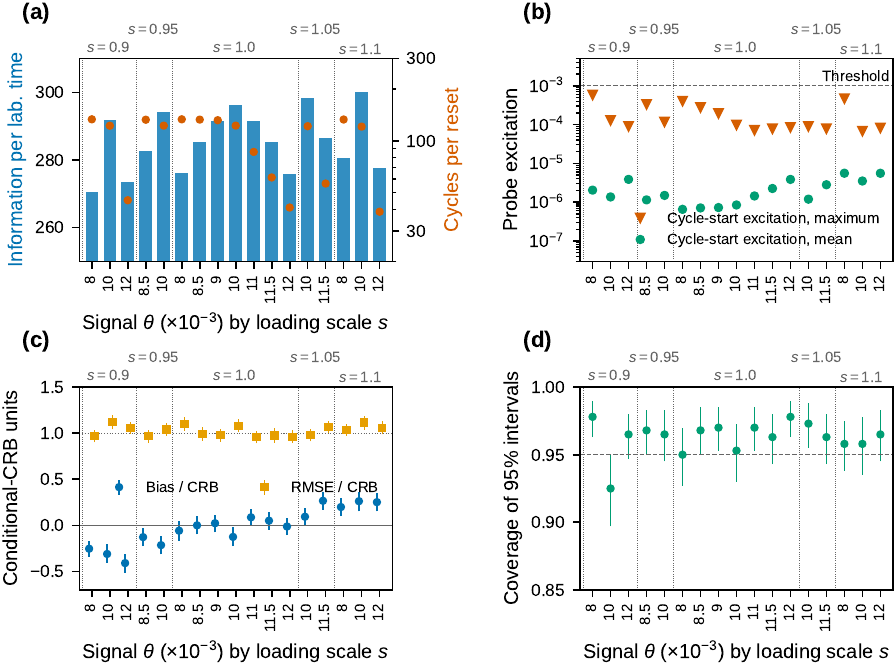}
\caption{Repeated operation of the baseline protocol at the seventeen simulated truths of Table~\ref{tab:settings}, 400 independent histories of 400 cycles each; points are grouped by the true loading scale $s$ and labelled by the true signal $\theta$. (a) Information per unit of laboratory time (bars, left axis) and cycles per reset including the three start-up resets (circles, right axis; 95\% bootstrap intervals over histories). (b) Mean (circles) and maximum (triangles) sensor excitation at the start of a cycle, after recovery; the dashed line is the controller's threshold. (c) Bias and root-mean-square error of the complete-record estimator after 400 cycles, in units of the conditional Cram\'er--Rao scale $1/\sqrt{400F}$, with bootstrap intervals; the loading scale is averaged over. (d) Empirical coverage of the nominal 95\% credible intervals, with bootstrap intervals.}
\label{fig:cycle}
\end{figure}

We now run the complete cycle in the open-system model at seventeen combinations of the true signal and the true loading scale (Table~\ref{tab:settings}); these truths lie in $\theta\in[0.008,0.012]$ and $s\in[0.9,1.1]$, and the statements below are about them, not about every point of the prior. Each point has 400 independent histories of 400 cycles, $2.72\times10^6$ cycles in all. The controller knows neither $\theta$ nor $s$; it evolves fifteen candidate sensor states, one for each pair of its signal and scale candidates, conditioned on the observed outcomes, and acts on the worst of them. Table~\ref{tab:ops} lists every operation and its assigned duration, and Fig.~\ref{fig:cycle} collects the results.

First, no history failed: none was aborted or halted, and the largest sensor excitation at the start of a cycle, after recovery, was $9.0\times10^{-4}$, with means between $6\times10^{-7}$ and $8\times10^{-6}$ [Fig.~\ref{fig:cycle}(b)]; the largest excitation immediately before a readout was $5.8\times10^{-4}$, and a single post-readout conditional estimate in the whole run ($1.35\times10^{-3}$) exceeded the threshold, which the recovery reset then removed. Second, resets are rare: between one per 39 and one per 133 cycles including the three start-up resets, the higher rates occurring at the larger signals [Fig.~\ref{fig:cycle}(a)]. Third, the information per unit of laboratory time, with every reset and interval counted, lies between 270 and 300 at all seventeen points.

Note that ``no event occurred'' carries its own uncertainty: with 400 histories per point, the zero counts bound the per-history event probability at $0.75\%$ pointwise ($1.45\%$ after a Bonferroni correction across the seventeen points), and these are per-history statements, since the controller's actions make the cycles of a history dependent.

We turn to estimation. The estimator is the Bayesian posterior of the complete record of helper outcomes, evaluated on the discrete grid of Table~\ref{tab:settings} with the conditional maps of the open-system model, with uniform priors on the signal and the loading scale and the scale averaged over; it also carries the controller's reset decisions. After 400 cycles its bias is at most $0.41$ of the conditional Cram\'er--Rao scale $1/\sqrt{400F}$, where $F$ is the per-cycle Fisher information at the true parameters, its root-mean-square error is $0.96$--$1.12$ of that scale, and the empirical coverage of the nominal 95\% credible intervals is $0.925$--$0.978$ [Fig.~\ref{fig:cycle}(c,d)]. Refining the grids changes these numbers within their bootstrap intervals at the three points checked. Note that the conditional scale is a reference, not an attainable bound for a biased finite-record estimator, and that it assumes the coupling known.

The record itself identifies the loading scale only weakly (nuisance-adjusted Fisher information \cite{Suzuki2020} for $s$ of $2.3\times10^{-7}$ per cycle at the nominal point): to leading order the readout depends on $\theta$ and $s$ only through the accumulated phase $\int\omega\,dt\simeq2\theta\!\int b(sJ)\,dt$, an approximate metrological symmetry of the kind analysed in Ref.~\cite{Mihailescu2026}. So a ten per cent miscalibration shows up as a bias of up to $0.4$ statistical standard deviations at 400 cycles, of the sign the local sensitivity predicts; at this rate the calibration bias would equal the statistical uncertainty after a few thousand cycles, an extrapolation beyond the tested length, after which an external calibration or a reference measurement is needed. A design study with known reference offsets, which shows that offsets of $\pm0.02$ and $\pm0.03$ about $\theta_c$ make the scale identifiable while retaining 80\% and 90\% of the signal information, is given in Appendix~\ref{app:cycle}; it is a local design prediction, and the paper does not claim self-calibration.
\section{Comparison with direct sensing on a common metric}
\label{sec:direct}

Does reading the sensor through a helper cost information compared with reading the sensor itself? To answer this we built two kinds of direct reference in the same open-system model, with the same bath and the same accounting of laboratory time, and we keep them apart because they were optimised differently. The first is a frontier: at each duration and noise setting the directly read sensor is prepared by a bounded five-unit transverse pulse, dressed by a local $\sx$ field ramped at the same slew limit to a strength $J_d\in\{0.5,1,1.5,2,3\}$ chosen for the best rate, held, unloaded and read along the best axis on its Bloch sphere with the same one per cent assignment error; its projective readout leaves it in a mixed state, so it is reset once per cycle. The second is a fixed instrument built on the baseline-duration member of that family ($J_d=1.5$), run as a complete repeated cycle with nondemolition verification of its readout (Table~\ref{tab:ops}); it inherits the dressing, preparation and readout axis of the first and adds the verification logic.

\begin{figure}
\centering\includegraphics[width=\textwidth]{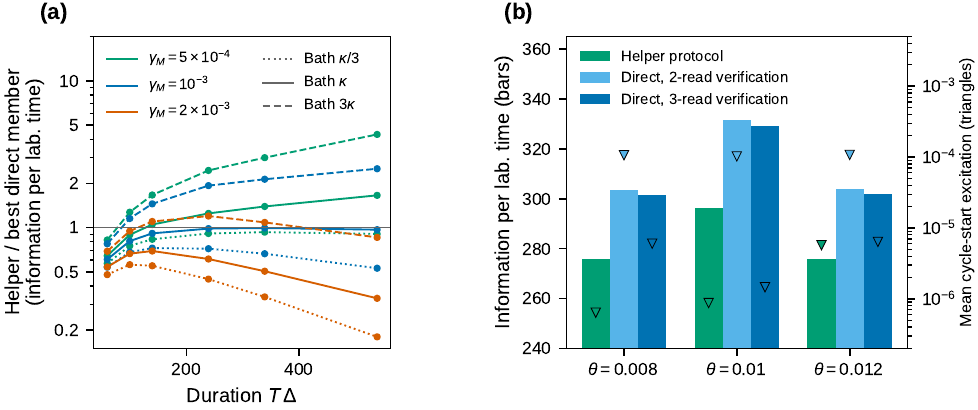}
\caption{Helper protocol against direct sensing. (a) Ratio of the information per unit of laboratory time collected by the helper protocol to that of the best member of the direct family, as a function of duration, for three bath strengths (line style) and three helper dephasing rates (colour); the direct dressing strength is chosen at each point. (b) Repeated-cycle comparison at the baseline duration ($T=138.5$) at three true signals ($s=1$): information per unit of laboratory time (bars) and mean sensor excitation at the start of a cycle (triangles, right axis) for the helper protocol and for the directly read sensor with blocks of two or three verification reads; 400 histories of 400 cycles, analysis rotation and conditional return ideal and instantaneous.}
\label{fig:direct}
\end{figure}

Figure~\ref{fig:direct}(a) maps the ratio of the two rates over the duration ($58$ to $539$), the bath strength ($\kappa/3$, $\kappa$, $3\kappa$) and the helper dephasing rate ($0.5$, $1$ and $2\times10^{-3}$). At the nominal noise the helper trails the best direct member by 39\% at the shortest duration and by $1$--$3\%$ at durations of 240 to 540; its rate is limited by the helper's own dephasing, which is why a quieter helper or a noisier sensor bath tips the balance, up to a factor of four in the helper's favour at long durations with a bath three times stronger and a helper twice quieter. The direct family is a one-parameter family of dressings, not all controls, so the map says where the helper is competitive, not that it is optimal.

A directly read sensor can also protect itself, by verifying its own readout \cite{Braginsky1980,Vaartjes2026}. In our instrument the sensor is rotated into the energy basis of the design point and read $m$ times; each read is the instrument $\mathcal M_y(\rho)=(1-e)\hat P_y\rho\hat P_y+e\hat P_{1-y}\rho\hat P_{1-y}$ for the reported outcome $y$, with $e=0.01$, followed by one time unit of bath evolution and measurement dephasing. A block is accepted when all $m$ reports agree and repeated otherwise, and the sensor is then returned by an instantaneous swap of the two energy states if the accepted report is ``excited'' and left alone if it is ``ground'' (Appendix~\ref{app:direct}). Note that a reported outcome does not leave a pure known state when assignment errors are present, which is why the cycle-start excitation of this instrument is not zero; device-level models of charge-sensor readout show the same feature \cite{Svastits2026}. The result, in Fig.~\ref{fig:direct}(b), is plain. With two-read verification the direct sensor collects $331.6$, $303.3$ and $303.8$ units of information per unit time at $\theta=0.010$, $0.008$ and $0.012$, against the helper protocol's $296.3$, $276.0$ and $275.8$, with a mean cycle-start excitation of $1.0$--$1.1\times10^{-4}$ (helper: $0.6$--$5.7\times10^{-6}$); with three reads the rates are $329.3$, $301.5$ and $301.9$ and the excitation $1.5$--$6.4\times10^{-6}$, i.e.\ a cycle-start excitation of the same order as the helper's at $9$--$11\%$ more information. No cycle-start excursion above $10^{-3}$ occurred. The direct rate depends weakly on the duration of a read (two-read rate from $336$ to $314$ as it goes from $0.1$ to $5$), and about 16 time units of unmodelled gate time per cycle would be needed to bring the three-read rate down to the helper's. At the duration of the echo sequence of the next section ($T=185.1$) the margin widens to $25$--$42\%$ (Appendix~\ref{app:direct}).

The selected direct reference therefore has the higher information rate. Its analysis rotation and conditional return are ideal and instantaneous, which favours it; its one-parameter dressing family and fixed preparation do not. The helper protocol's possible motivation is of a different kind: the sensor is never read, so whatever a direct readout costs in the device at hand, beyond what the model includes, is avoided. With a hypothetical per-read damage probability the two break even when a single direct read is about $0.4$ times as likely to damage the sensor as a whole helper cycle is (Appendix~\ref{app:direct}); whether a device meets that condition is a property of the device, for which we have no model.
\section{A helper echo that preserves the band}
\label{sec:echo}

A plain Ramsey sequence on the helper cannot tell the signal from a slow offset of the helper's own frequency. If the helper carries a term $\beta\tz$ with $\beta$ constant over a sequence, the branch-energy difference becomes $\omega-2\beta$, and the offset is read as a signal. For a Gaussian offset of width $\sigma$ shared by a whole record, the average contrast of a single sequence is $W=\exp(-2\sigma^2T^2)$, which is $2\times10^{-7}$ at $\sigma=0.02$ for the baseline pulse: the averaged single-sequence signal is strongly suppressed. The textbook remedy is an echo \cite{Hahn1950,Viola1998}: flip the helper half-way and let the second half undo the offset phase. However, a flip of the helper at fixed coupling is not a symmetry of the dressed band: it exchanges the two branches, and in the model it drives the sensor out of the band with a leakage of $0.97$.

The flip must therefore be performed at $J=0$. With the coupling off, the band projector is $\ket{g}\bra{g}\otimes\openone$, which commutes with every operator on the helper alone, so a helper rotation at zero coupling generates no leakage at all; this is exact, not a numerical finding. The sequence is thus: load, hold, unload, flip the helper with a smooth $\pi$ pulse $\Omega(t)\tx$ of eight time units at $J=0$, and load and hold again. If the second half uses the same coupling as the first, the flip reverses the signal phase along with the offset phase and the sequence is blind to a constant signal; reversing the sign of the coupling in the second half reverses the branch assignment instead, so that the signal phase adds while the offset phase cancels (Appendix~\ref{app:pulse}). This is the \emph{signed echo}, with holds of 50 at $+J$ and $-J$ around the flip ($T=185.1$) [Fig.~\ref{fig:scheme}(c)]. It needs a sign-reversible coupling; an alternative that keeps the sign and breaks the symmetry with a reference interval is less efficient (Appendix~\ref{app:numerics}). Note that the cancellation is exact only for an instantaneous flip and an ideal symmetry; for the finite, detuned pulse used here the suppression is strong but numerical.

\subsection{Single sequences}

\begin{figure}
\centering\includegraphics[width=\textwidth]{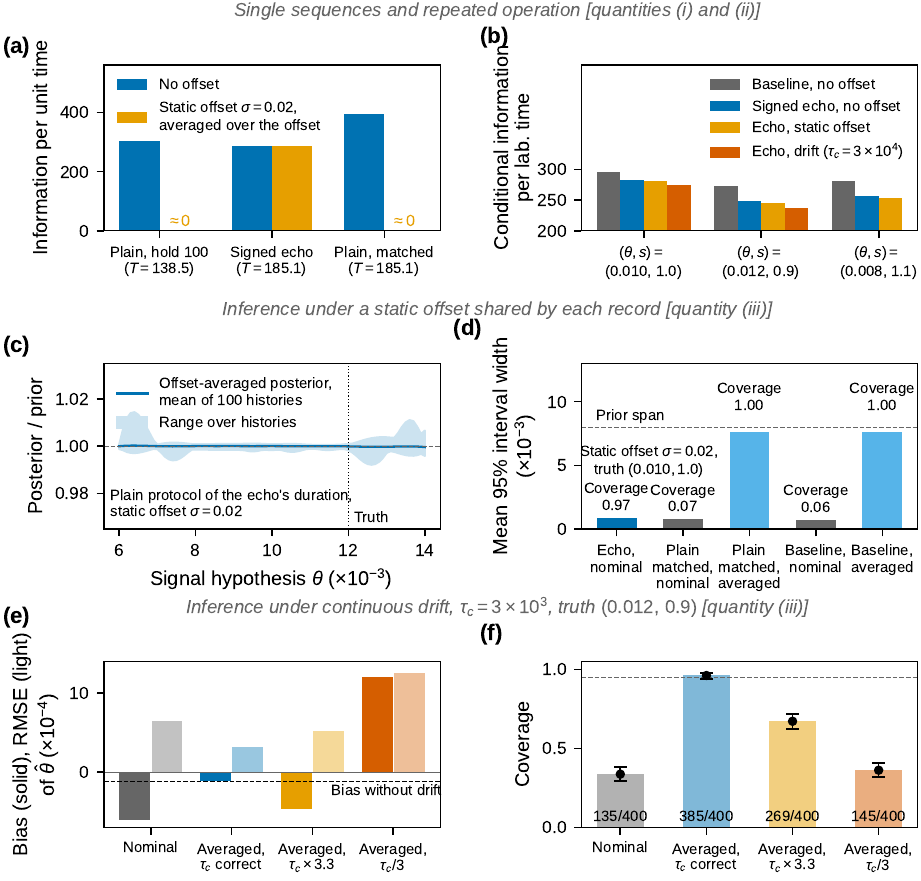}
\caption{The signed echo under helper offsets and drift. Top row, single sequences and repeated operation without inference: (a) information per unit time of single sequences without an offset and under a static Gaussian offset of width $\sigma=0.02$ averaged at the level of the sequence probabilities [quantity (i) of the text]; (b) conditional information per unit of laboratory time through the complete cycle [quantity (ii)] at three truths for the baseline, the echo, the echo under a static offset drawn once per history, and, at two truths, the echo under continuous drift with correlation time $3\times10^4$; 400 histories of 400 cycles. Middle row, inference under a static offset shared by each record [quantity (iii)]: (c) posterior of the plain protocol of the echo's duration at the truth $(0.012,\,0.9)$, averaged over the shared offset with the exact 401-node rule, relative to the uniform prior (mean and range over 100 histories); (d) mean width of the 95\% credible intervals and their coverage at the truth $(0.010,\,1.0)$ for the echo with the nominal estimator (400 histories) and for the plain protocols with the nominal and the offset-averaged estimators (100 histories); the dashed line is the span of the prior. Bottom row, inference under continuous drift with $\tau_c=3\times10^3$ at the truth $(0.012,\,0.9)$, 400 histories: (e) bias and root-mean-square error of the signal estimate for the nominal estimator and for the drift-averaged estimator with the correct, a $3.3$ times too long and a $3$ times too short assumed correlation time; the dashed line is the bias without drift; (f) coverage of the same estimators with 95\% Wilson intervals, counts out of 400.}
\label{fig:echo}
\end{figure}

In the open-system model the signed echo collects $53\,190$ units of information per sequence, which is $72.8\%$ of what a plain pulse of the same duration collects ($73\,044$, hold $146.5$) and $95\%$ of the baseline pulse's information per unit time [Fig.~\ref{fig:echo}(a)]. The cost of about 27\% has a simple budget. Relative to the plain pulse, the echo exchanges $46.5$ time units of hold at the full phase gain for two more ramps, which do accumulate signal phase but at $56\%$ of the full gain on average, and for eight units of flip at zero coupling, where no signal phase accrues; a phase-only estimate with a common contrast factor then predicts an information ratio of $0.727$, close to the full open-system value of $0.728$. Thus the reduced integrated phase gain explains most of the observed loss; the full simulation includes the remaining finite-pulse and dissipative effects, which need not be identical for the two sequences. Under a static Gaussian offset of width $\sigma=0.02$, averaged at the level of the single-sequence probabilities, the signed echo retains $99.2\%$ of its information ($95.4\%$ at $\sigma=0.05$). Its peak band leakage over the sequence is $3.5\times10^{-7}$ for a fresh input, a $\pm2\%$ error in the flip amplitude costs $0.2\%$ of the information, and the helper's dephasing sets the scale of what remains ($74\,070$, $53\,190$ and $36\,817$ per sequence at $\gamma_M=10^{-4}$, $10^{-3}$ and $2\times10^{-3}$). The quadrature rules behind the offset averages are validated in Appendix~\ref{app:numerics}.

\subsection{Repeated operation and a static shared offset}

Three information quantities appear below and are kept apart: (i) the Fisher information of the offset-averaged single-sequence probability [Fig.~\ref{fig:echo}(a)]; (ii) the conditional record information, accumulated along the true noise path with the nuisance parameters fixed, an upper reference for the usable information of a record [Fig.~\ref{fig:echo}(b)]; and (iii) the error, width and coverage of an actual estimator after the nuisance is averaged over its prior \cite{ZhangSuzuki2025}, the operational quantities we compare.

Through the complete cycle, without any offset, the signed echo keeps $91$--$95\%$ of the baseline's conditional information per unit of laboratory time at the three truths tested: $282.7$, $248.1$ and $256.5$ against $296.3$, $273.5$ and $280.7$ at $(\theta,s)=(0.010,1.0)$, $(0.012,0.9)$ and $(0.008,1.1)$ [Fig.~\ref{fig:echo}(b)]. It resets once per 110, 30 and 133 cycles, shows no abort or halt, and its estimator has a root-mean-square error of $1.07$--$1.17$ conditional-CRB units with coverage $0.948$--$0.968$. With a static offset $\sigma=0.02$ drawn once per history the picture hardly changes ($280.5$, $245.8$ and $254.2$ per unit time, errors of $1.07$--$1.18$ units, coverage $0.943$--$0.970$).

For a static offset shared across a record, the loss of the averaged single-sequence contrast does not by itself establish the loss of all record information, so we evaluate the inference on the full tested records with the shared offset averaged over. The plain protocol of the echo's duration keeps a large conditional information under the offset ($233$ per unit time), but its tested records produce negligible narrowing of the prior. Its nominal estimator, which assumes no offset, has an error of $11$--$14$ conditional-CRB units and a coverage of $6$--$8\%$; the offset-averaged estimator returns essentially the prior [Fig.~\ref{fig:echo}(c,d)]. 

The latter rests on an exact integration over the offset (a 401-node rule, Appendix~\ref{app:drift}), so that only the $\theta$ and scale grids remain discrete. The posterior's total-variation distance from the prior is $4\times10^{-5}$ to $2\times10^{-3}$ over 100 histories at each of two truths, its mean is $0.010$ at both tested truths, and its 95\% width is the prior's own central width of $7.6\times10^{-3}$; the baseline pulse behaves in the same way.

Under the stated prior, settings and record length, the plain protocol thus produces little posterior narrowing whereas the signed echo retains useful estimation accuracy. This is a finite-record result within the specified model and for the implemented estimators, not an impossibility statement for every record length, prior or estimation strategy.

\subsection{Continuous drift}

A static offset is the easy case for an echo. To test a drifting one we let $\beta(t)$ be a stationary Ornstein--Uhlenbeck process \cite{Gardiner2009} with standard deviation $\sigma=0.02$ and correlation time $\tau_c$, advanced through every charged interval of every history; outside the flip only its integral over each segment matters, and during the eight-unit flip it is frozen at its value at the pulse centre, a midpoint approximation quantified below (Appendix~\ref{app:drift}). The controller and the nominal estimator know nothing of the drift. We tested $\tau_c=3\times10^4$ (about 160 sequence lengths) and $3\times10^3$ (about 16); the diffusion of the offset within a sequence, which the ideal instantaneous echo considered here does not cancel, costs that echo $5.5\%$ and $42\%$ of its phase information at the two correlation times.

At $\tau_c=3\times10^4$ the echo's conditional information falls by $2.8\%$ at the centre and $4.4\%$ at the off-centre truth $(0.012,0.9)$, and at $\tau_c=3\times10^3$ by $22\%$ at both. The nominal estimator, which assumes the no-drift contrast, is biased. At the centre of the prior its coverage and width are unchanged, but at the off-centre truth its coverage drops to $0.907$ ($363/400$) at the slower drift and to $0.338$ ($135/400$) at the faster one, with a bias of $-6.1\times10^{-4}$, five times the no-drift value [Fig.~\ref{fig:echo}(e,f)]. The mechanism is shrinkage toward the calibration point: a likelihood that assumes the no-drift visibility reads the drift-reduced contrast $W$ as a smaller signal phase, so that in the small-phase limit $\hat\theta-\theta_c\simeq W(\theta-\theta_c)$. With the contrast of the actual sequence under the drift, $W=0.972$ and $0.759$, this predicts additional biases of $-5.5\times10^{-5}$ and $-4.8\times10^{-4}$; the observed values are $-7.7\times10^{-5}$ and $-4.9\times10^{-4}$. At the centre the truth is the calibration point and the shrinkage is invisible.

The remedy is to put the contrast into the likelihood. We average the single-cycle channel over the stationary law of the offset's segment integrals and its value at the flip, with the exact covariances of the process, and apply this averaged likelihood cycle by cycle as if the cycles were independent: a composite likelihood \cite{Lindsay1988,Varin2011} that carries the drift's contrast loss but not the correlation of the drift across cycles. On the same records it removes the additional bias: at $\tau_c=3\times10^3$ the bias returns to $-1.1\times10^{-4}$, the error to $1.22$ conditional-CRB units and the coverage to $0.963$ ($385/400$, Wilson interval $[0.939,\,0.977]$), with intervals widened to $1.3\times10^{-3}$ as the reduced contrast requires, and at $\tau_c=3\times10^4$ the coverage is $0.9325$ ($373/400$, $[0.904,\,0.953]$).

This repair is conditional on knowing the drift law. The averaged estimator was given the true $\sigma$ and $\tau_c$. With the correlation time assumed $3.3$ times too long it leaves three quarters of the bias in place ($-4.7\times10^{-4}$, coverage $269/400$); assumed three times too short, it over-corrects and pushes the estimate past the truth, with a bias of $+1.2\times10^{-3}$, an error of $1.26\times10^{-3}$ ($3.9$ times the correctly specified value and twice the nominal estimator's) and coverage $145/400$ [Fig.~\ref{fig:echo}(e,f)]. Both follow the shrinkage rule with the assumed contrast in place of the true one, $\hat\theta-\theta_c\simeq(W_{\rm true}/W_{\rm assumed})(\theta-\theta_c)$. The estimator therefore requires adequate noise characterisation; learning $\sigma$ and $\tau_c$ from the record is not implemented (real-time estimation of slowly fluctuating Hamiltonian parameters has been demonstrated for spin qubits \cite{Berritta2024} and is the natural route), and the three mismatch cases do not establish a tolerance range.

Three limitations remain. The intervals of the composite likelihood are not certified: weak correlations of the residual phase noise persist over many lags (the variance of the 400-cycle phase average is $0.53$ and $0.30$ of its independent-cycle value in an idealised diagnostic), the closeness of the composite likelihood to the full one has not been established, and the empirical coverages above stand on their own. The midpoint approximation during the flip was checked as a single-cycle diagnostic only (at the faster drift and the off-centre truth, a mean readout-probability discrepancy of $(2.40\pm0.97)\times10^{-4}$ and a root-mean-square discrepancy of $0.93\%$ of the drift's per-cycle effect; Appendix~\ref{app:drift}); its accumulated effect on the full-record estimator has not been measured. Protection is maintained in the sampled runs but its statistics change: no excitation above $10^{-3}$ was observed at any drift setting, while at the off-centre truth the reset rate moved from one per 30 cycles to one per 32 and one per 46 (Appendix~\ref{app:drift}).
\section{Feasibility}
\label{sec:device}

The model is generic and we have not tied it to a device. Table~\ref{tab:req} separates what was tested from what a device would have to provide; only the coupling threshold was derived, the other entries are tested values. In units of the sensor splitting $\Delta$, the coupling must be of the order of the splitting or larger ($J\ge1.03\Delta$ keeps the gain $b$ above $0.9$ of its maximum), switchable to zero within a cycle and, for the signed echo, sign-reversible; the counterdiabatic terms require a joint $\sy\tz$ drive of amplitude up to $0.07\Delta$ and a local $\sy$ drive below $0.002\Delta$. The helper's Markovian dephasing sets the information per sequence (Sec.~\ref{sec:echo}), so $\gamma_M$ is the parameter a device must match first, and signal-like noise on the sensor, i.e.\ fluctuations of $\theta$ itself, is filtered by nothing in the protocol.

\begin{table}
\caption{Tested parameters of the protocol and the requirements that follow from them, in units of the sensor splitting $\Delta$ (times in units of $1/\Delta$). ``Tested'' entries are operating values at which the results hold; only the coupling threshold was derived.}
\label{tab:req}
\footnotesize
\centering\begin{tabular}{p{0.25\textwidth}p{0.32\textwidth}p{0.28\textwidth}}\toprule
quantity & tested & requirement \\
\midrule
coupling $J_{\max}$ & $3$, calibrated to $\pm10\%$ & $\ge1.03$ for $b\ge0.9$; switchable to $0$ \\
coupling sign & reversed in the echo & reversible (signed echo only) \\
counterdiabatic drives & $\sy\tz$ to $0.07$; $\sy$ to $0.002$ & same waveforms \\
ramp, hold & $19.3$; $100$ (echo $2\times50$) & as tested \\
helper flip & $\tx$ $\pi$ pulse over $8$ & at $J=0$ \\
helper dephasing $\gamma_M$ & $10^{-3}$ ($10^{-4}$, $2\times10^{-3}$) & tested values only \\
helper offset & Gaussian, $\sigma=0.02$, $0.05$ & tested values only \\
sensor bath & $\kappa=10^{-3}$, $k_BT_b=0.1$ & tested values only \\
sensor reset & rate $10$, once per $30$--$133$ cycles & engineered damping \\
helper readout & binary, error $0.01$ & as tested \\
\bottomrule
\end{tabular}
\end{table}

For a concrete scale, consider two charge qubits in Si/SiGe double quantum dots \cite{Burkard2023}: the sensor a double dot at zero detuning, so that its tunnel coupling $t$ sets $\Delta=2t$ and a detuning signal is transverse, the helper a second double dot held at large detuning, so that its position is its energy eigenbasis $\tz$. The measured inter-double-dot capacitive interaction of Ref.~\cite{Neyens2019}, $g/h=20.9\,$GHz, then takes the form $J\sx\tz$ with $J=g/4$, once the accompanying local detuning terms are compensated. At the measured setting the sensor tunnel coupling was $t/h=5.8\,$GHz, i.e.\ $\Delta/h=11.6\,$GHz and $J/\Delta\simeq0.45$, below the threshold; a hypothetical retuning to $t/h=1\,$GHz with the same $g$ would give $J/\Delta\simeq2.6$, above the $b\ge0.9$ threshold though short of the tested $J/\Delta=3$, and this combination has not been demonstrated jointly. The intradot tunnel couplings reported there are a different quantity and do not by themselves supply the interaction. 

That coupling is set by gate geometry and tuned slowly, however, whereas the protocol switches $J$ off and, for the echo, reverses it within one cycle, and the joint $\sy\tz$ correction has no demonstrated counterpart. 

The temperature requirement $k_BT_b\le0.1\Delta$ reads $T_b\le14.4\,$mK at $\Delta/h=3\,$GHz and $24.0\,$mK at $5\,$GHz, within reach of dilution refrigerators though not of the $155\,$mK electron temperature of that device; charge-qubit dephasing times of nanoseconds \cite{Shi2013} are the more serious obstacle (charge noise below $2\,\mu$eV$/\sqrt{\rm Hz}$ is now reported in industrially fabricated Si/SiGe devices \cite{Koch2025}, though that figure does not by itself give a charge-qubit coherence time), since $\gamma_M\lesssim10^{-3}\Delta/\hbar$ asks for a helper coherence time $T_2\gtrsim10^3\,\hbar/\Delta$, i.e.\ $53\,$ns at $\Delta/h=3\,$GHz and $32\,$ns at $5\,$GHz (about 160 oscillation periods). Spin--charge hybrid or singlet--triplet helpers, and superconducting circuits with tunable couplers \cite{Ding2023,Coupler2024}, are the candidates we would examine next; tunable couplers demonstrate the coupling control, not the full regime, and the partly refocusable inhomogeneous dephasing of such helpers is what the signed echo would rely on, while the residual white-noise dephasing rate must still satisfy $\gamma_M\lesssim10^{-3}\Delta/\hbar$. No compatible device has been demonstrated, and a device proposal would have to meet the interaction form, the coupling strength, the reset, the readout and the residual-noise conditions together.
\section{Conclusion}
\label{sec:conclusion}

The protocol encodes a transverse signal in the phase of a helper qubit while keeping leakage out of a selected dressed band small. In the tested open-system model, calibrated loading and unloading return the sensor close to its idle ground state and support repeated operation with an uncertain coupling and infrequent resets, and the baseline and static-offset echo runs give credible-interval coverages close to their nominal values at the tested truths. A helper flip at zero coupling, combined with a reversal of the coupling sign, suppresses a static offset shared across a record while retaining the signal, at a cost of about 27\% of the no-offset information of a plain sequence of equal duration. Under slowly varying noise an estimator that ignores the contrast loss is biased toward the calibration point, and averaging over the noise removes that bias only when the assumed noise law is adequate. The selected direct sensor with nondemolition verification keeps the higher information rate under its ideal controls.

These results establish a protocol and its trade-offs within a weak-coupling driven master-equation model in which measurement and reset interventions are applied between pulses; no compatible device and no practical advantage is demonstrated. The record does not reliably calibrate the coupling, and the composite-likelihood intervals under drift carry no coverage guarantee. The next steps are a device-specific model of readout disturbance, an estimator that accounts for the noise correlations between cycles, and a calibration run with uncertain reference signals. Together they would decide whether avoiding direct readout outweighs its information-rate cost in a device.

\backmatter

\bmhead{Acknowledgements}
The author thanks the Kurdistan Institution for Strategic Studies and Scientific Research for financial support.

\section*{Statements and Declarations}

\bmhead{Funding}
This work was supported by the Kurdistan Institution for Strategic Studies and Scientific Research.

\bmhead{Competing interests}
The author declares no competing interests.

\bmhead{Data and code availability}
The simulation code, the stored channel caches, the run scripts with their fixed seeds, the logs from which every number in this paper is taken, and the script that generates the figures from them are available in the Zenodo record \url{https://doi.org/10.5281/zenodo.23796406}. The record's documentation describes the directory layout, the software versions, the build commands and a manifest that links each figure and each quoted number to its source files.

\bmhead{Author contributions}
S.A. is the sole author: conception, theory, numerical simulations, analysis and writing.

\bmhead{Use of generative AI}
Generative AI tools (Claude, Anthropic; Claude Fable 5.1 models) were used under the author's direction to assist with the numerical codes and the figures; all results were verified by the author, who takes full responsibility for the manuscript.

\begin{appendices}
\setcounter{figure}{5}\setcounter{table}{3}
\gdef\thefigure{\arabic{figure}}\gdef\thetable{\arabic{table}}

\section{Proof and numerical checks of the single-sensor bound}
\label{app:bound}

Let $\ket{\psi_t}$ be the state of the sensor under $\hat H(t)=\hat H_{\rm ref}(t)+u_y(t)\sy$, where the controls $u_x(t)$ and $u_y(t)$ are fixed functions chosen without knowledge of $\theta$ ($u_x$ bounded and of bounded variation, $u_y$ bounded and integrable, which covers piecewise smooth waveforms and, as a limit, sudden rotations; the integration by parts below is then to be read in the Stieltjes sense), starting from the ground state $\ket{g_0}$ of $\hat H_{\rm ref}(0)$, which depends on $\theta$. Writing $U(t)$ for the propagator, the generator of $\theta$ translations on the final state is $\hat G=\int_0^TU^\dagger(t)\sx U(t)\,dt$, and the quantum Fisher information is $F_Q=4\|\Pi_\perp\ket{\phi}\|^2$ with $\ket{\phi}=\ket{\partial_\theta g_0}-\ii\hat G\ket{g_0}$ and $\Pi_\perp$ the projector orthogonal to $\ket{g_0}$. The first term is bounded by first-order perturbation theory, $\|\Pi_\perp\ket{\partial_\theta g_0}\|\le1/\Delta$, since the gap of $\hat H_{\rm ref}$ is at least $\Delta$. The second term is the standard deviation of $\hat G$ in $\ket{g_0}$, which is a seminorm and therefore obeys the triangle inequality and is bounded by the operator norm.

Decompose $\sx$ in the instantaneous eigenbasis of $\hat H_{\rm ref}(t)$ as $\sx=V_{gg}\openone+(V_{ee}-V_{gg})\hat P_e(t)+\hat V_o(t)$, with $\hat P_e$ the excited projector of $\hat H_{\rm ref}(t)$ and $\hat V_o$ the off-diagonal part. The identity term contributes nothing to the standard deviation. The off-diagonal part can be written as a commutator, $\hat V_o=\ii[\hat H_{\rm ref},\hat B]$ with $\hat B=\Delta\sy/[\Delta^2+4(\theta+u_x)^2]$, as one verifies directly; because $[u_y\sy,\hat B]=0$ this also equals $\ii[\hat H,\hat B]$, and integration by parts gives $\int_0^TU^\dagger\hat V_oU\,dt=[U^\dagger\hat BU]_0^T-\int_0^TU^\dagger\dot{\hat B}U\,dt$. The standard deviation of the first bracket is at most $2\|\hat B\|_{\max}=2/\Delta$ and that of the integral at most $A=\int_0^T\|\dot{\hat B}\|\,dt$, with $\|\dot{\hat B}\|=|d(\Delta/\Omega^2)/dt|$ and $\Omega^2=\Delta^2+4(\theta+u_x)^2$. The remaining term is bounded by $\int_0^T|V_{ee}-V_{gg}|\,{\rm SD}_{\psi_t}(\hat P_e)\,dt\le2\int_0^T\sqrt{\epsilon_t(1-\epsilon_t)}\,dt\le2\sqrt{\epsilon}\,T$, using $|V_{ee}-V_{gg}|\le2$, where $\epsilon_t=\bra{\psi_t}\hat P_e(t)\ket{\psi_t}$ is measured against $\hat H_{\rm ref}$ and not against the full field. Collecting terms, ${\rm SD}(\hat G)\le2/\Delta+A+2\sqrt{\epsilon}\,T$ and Eq.~(\ref{eq:bound}) follows. $A$ is the total variation of $\Delta/\Omega^2$, at most $1/\Delta$ per monotone excursion of the gap; a signed dressing that crosses $\theta+u_x=0$ passes through the maximum $1/\Delta$ of $\Delta/\Omega^2$, and a monotone crossing contributes $2/\Delta-\Delta/\Omega^2_{\rm start}-\Delta/\Omega^2_{\rm end}$, up to $2/\Delta$, which is why the $2/\Delta$ simplification is stated for a dressing of fixed sign; the bound itself retains the actual $A$. With a $\theta$-independent preparation the first term is absent and the constant is $2/\Delta$; in that case, and without any excitation budget, one recovers the familiar $F_Q\le4T^2$ from ${\rm SD}(\sx)\le1$. With the $\theta$-dependent ground-state preparation the constant $3/\Delta$ remains, and some information is present even at $T=0$.

The numerical checks use these definitions: $\epsilon_t$ from the excited projector of $\hat H_{\rm ref}(t)$ with $u_y$ excluded, and $A$ as the total variation of $\Delta/\Omega^2$. The control family is six sine harmonics per axis with amplitudes limited to 3 by a $\tanh$ saturation, at $T=27.7$ and $58.5$.

Forty random members of this family reach at most $0.057$ and $0.13$ of the bound when both controls are used and $0.40$ and $0.15$ when only $u_x$ is used. A penalised Powell search over both controls, with four restarts, under $\epsilon_t\le\epsilon$ finds best values of $4.2$, $5.8$ and $7.9$ at $T=27.7$ and $4.7$, $7.3$ and $18.0$ at $T=58.5$ for $\epsilon=10^{-4}$, $10^{-3}$ and $2\times10^{-3}$, against bounds of $51$, $94$, $137$ and $72$, $186$, $321$. These are the best values found in that family by a local search, not global optima.

The kick-and-hold construction lies outside the smooth family. It prepares the sensor at the design signal $\theta_c$ in the dressed basis at $u_x=3$, rotates it suddenly to excitation $\epsilon$ and holds; the preparation is signal-independent, so its constant is $2/\Delta$. Its exact information is $1.13$ and $11.8$ at $T=27.7$ and $5.15$ and $52.7$ at $T=58.5$ for $\epsilon=10^{-4}$ and $10^{-3}$, against the leading long-time term $16\epsilon(1-\epsilon)b^2T^2=1.20$, $11.9$, $5.34$ and $53.3$ with $b=0.986$ at $u_x=3$; the leading term is approached from below at finite time. Because the kick angle is set at $\theta_c$, the excitation against the dressed basis of the true signal varies between $0.993\times10^{-3}$ and $1.007\times10^{-3}$ over the prior, so a budget of exactly $\epsilon$ at every truth requires the kick to be set slightly below it.

\section{Loading pulse, band projector and the echo symmetry}
\label{app:pulse}

The ramps follow $J(t)=\tfrac12\tan[\varphi_{\max}\,f(t/\tau_r)]$ with $f(u)=u-\sin(2\pi u)/2\pi$ and $\varphi_{\max}=\arctan(2J_{\max})$, so that the mixing angle of a $\theta=0$ branch advances smoothly from $0$ to $\varphi_{\max}$; the unloading ramp is the mirror image. The counterdiabatic coefficients are $c_0+c_Z=\dot J/[1+4(J+\theta_c)^2]$ and $c_0-c_Z=-\dot J/[1+4(J-\theta_c)^2]$, i.e.\ half the rates of change of the mixing angles $\varphi_\pm=\arctan[2(\theta_c\pm J)]$ of the two branches at the design signal $\theta_c$ and the nominal coupling; their peak magnitude is $\arctan(2J_{\max})/\tau_r=0.073$, almost entirely in the joint $\sy\tz$ term. The commanded $J$ in these coefficients is the nominal one, so a loading-scale error $s\neq1$ leaves a residual nonadiabatic coupling, which is the origin of the leakage of $3$--$4\times10^{-5}$ at $s=0.9$ and $1.1$ [the former is shown in Fig.~\ref{fig:scheme}(e)]. The reference band projector is built from the branch Hamiltonians $\hat H_\pm=\tfrac12\sz+(\theta\pm sJ)\sx$ at the true signal and the true scale, without the counterdiabatic terms; with the $\sy$ terms included the reference branch ground state overlaps the instantaneous excited state by up to $3.8\times10^{-3}$ during the ramps, which is why both the bound and the leakage figures use the reference convention. The ramps accumulate signal phase: the slope $d\Phi/d\theta$ of the ramp-integrated branch phase $\Phi=\int\omega(\theta,J(t))\,dt$ is $21.4$ per ramp against $38.0$ for the same time at full gain, i.e.\ $56\%$, and $240.1$ for the whole baseline pulse.

At $J=0$ the two branch Hamiltonians coincide and the band projector reduces to $\ket{g}\bra{g}\otimes\openone$, which commutes with every operator of the form $\openone\otimes\hat O$; a helper rotation at zero coupling therefore maps the band to itself exactly, whatever its waveform. The echo symmetry can be seen on the ideal sequence. Write the helper state as $\alpha\ket{\uparrow}+\gamma\ket{\downarrow}$ and let the sensor follow the branch ground states adiabatically. In the first half the $\ket{\uparrow}$ component accumulates the phase $\int(E_{g,+}+\beta)\,dt$ and the $\ket{\downarrow}$ component $\int(E_{g,-}-\beta)\,dt$, so their relative phase grows at $\omega-2\beta$. An instantaneous flip exchanges the helper labels. If the second half uses the same coupling, the component now labelled $\ket{\downarrow}$ sees $E_{g,-}$ and the offset $-\beta$: both the signal phase $\int\omega\,dt$ and the offset phase $2\int\beta\,dt$ change sign and cancel, and the sequence is blind to a constant signal. If instead the coupling sign is reversed, the branch Hamiltonians are exchanged along with the labels, each component continues to see the same sensor energy, the signal phase adds to $2\int_0^{T/2}\omega\,dt$, and only the offset phase cancels. For the finite eight-unit flip with the sensor unloaded, the same bookkeeping holds up to the detuning of the flip and the ramps, which is why the suppression is numerical ($99.2\%$ retained at $\sigma=0.02$) rather than exact. The helper flip is $\Omega(t)=(\pi/8)\sin^2(\pi t/8)$ over eight time units.

\section{Open-system construction, the positive echo and the offset averages}
\label{app:numerics}

The jump operator of the universal Lindblad equation (ULE) is $\hat L(t)=\int ds\,g(s)\,U(t,t-s)\hat XU^\dagger(t,t-s)$ with $\hat X=\sz\otimes\openone$ and $g(s)$ the Fourier transform of $\sqrt{\gamma(\omega)}$, where $\gamma(\omega)=\kappa\,\omega\,\e^{-|\omega|/\omega_c}/(1-\e^{-\omega/k_BT_b})$ is the emission-convention bath spectrum; the master equation is $\dot\rho=-\ii[\hat H+\hat H_{LS},\rho]+\hat L\rho\hat L^\dagger-\tfrac12\{\hat L^\dagger\hat L,\rho\}$ plus the helper term $\gamma_M(\tz\rho\tz-\rho)/2$, with the Lamb shift $\hat H_{LS}$ built from the same kernel \cite{Nathan2020}. The driven propagator and the memory integrals are evaluated on a uniform grid of step $0.005$ with a memory window of $\pm150$ time units; the Lamb-shift window is $\pm20$, evaluated every ten steps. The accuracy parameter $\Gamma\tau_B=4(\int|g|)^2\int|sg|/\int|g|$ is $0.016$. Table~\ref{tab:conv} shows the convergence of the quantities that the protection claim rests on, computed with the production generator (ULE with Lamb shift on the periodic timeline) for the hold-100 pulse at $\theta=0.010$: the step is halved and the memory window is extended from $\pm150$ to $\pm250$ (the mass of $|g|$ outside the $\pm150$ window is $1.1\times10^{-5}$ of its total). The end-point excitation, which equals the band leakage at the end of the pulse, moves by $0.07\%$ under the step halving and is unchanged at the reported precision under the window extension; the maximum band leakage at $s=0.9$, which is set by the ramps, is unchanged at four digits.

\begin{table}
\caption{Convergence of the small leakage and end-point values under the propagation step $\delta t$ and the memory window $\pm S$ (hold 100, $\theta=0.010$, fresh input; production setting in the first row). At $s=1$ the maximum band leakage over the pulse occurs at its end and equals the end-point excitation; at $s=0.9$ the maximum occurs in the ramps. $F_1$ is the information per pulse at $s=1$.}
\label{tab:conv}
\footnotesize
\centering\begin{tabular}{cccccc}\toprule
$\delta t$ & $S$ & \begin{tabular}[c]{@{}c@{}}leakage, $s=1$\\($\times10^{-7}$)\end{tabular} & \begin{tabular}[c]{@{}c@{}}maximum leakage,\\$s=0.9$ ($\times10^{-5}$)\end{tabular} & \begin{tabular}[c]{@{}c@{}}end excitation,\\$s=0.9$ ($\times10^{-7}$)\end{tabular} & $F_1$ \\
\midrule
$0.005$ & $150$ & $1.6702$ & $3.5329$ & $2.6610$ & $41950.02$ \\
$0.0025$ & $150$ & $1.6691$ & $3.5329$ & $2.6603$ & $41950.04$ \\
$0.005$ & $250$ & $1.6702$ & $3.5329$ & $2.6610$ & $41950.02$ \\
$0.0025$ & $250$ & $1.6691$ & $3.5329$ & $2.6603$ & $41950.04$ \\
\bottomrule
\end{tabular}
\end{table}

The time-convolutionless second-order (TCL2) benchmark uses the full driven propagator and the exact second-order memory kernel without secular or Markov approximations; it agrees with the ULE information per pulse to $0.011\%$ at hold 100 and $0.003\%$ at hold 20, and a phenomenological model with a smooth memory kernel gives the same qualitative picture.

For the sequence channels the flip window is propagated with the offset $\beta\tz$ in the Hamiltonian and the $\beta=0$ jump operator and Lamb shift, and interpolated in $\beta$ by a cubic spline on 97 points (interpolation errors of $1.3$--$5.9\times10^{-9}$ at four midpoints, a check made when the tables were built and recorded in the code's documentation rather than in a log); outside the flip the offset enters as the exact conjugation ${\rm Phase}(x)=\exp(-\ii x\,{\rm ad}_{\tz})$, so the full sequence channel at a static offset is $S(\beta)={\rm Phase}(\beta T_{\rm after})S_{\rm after}S_{\rm flip}(\beta){\rm Phase}(\beta T_{\rm before})S_{\rm before}$.

Besides the signed echo, we tested a \emph{positive echo} that keeps the sign of the coupling and breaks the symmetry between the halves with a reference interval instead ($T=285.1$). It collects $31\,370$ units of information per sequence, only $20.6\%$ of its matched plain pulse ($152\,016$), because the reference interval is dead time; it retains $99.2\%$ under a static offset of width $\sigma=0.02$. It needs only a helper drive, not a sign-reversible coupling, but it is the less efficient instrument and is not carried further.

The single-sequence results under a static offset average the sequence probabilities over the Gaussian offset, and this average needs care. For the plain pulse the offset commutes with the whole generator and the average is exact. For the echoes the conditional state depends on the offset through the flip, and Gauss--Hermite quadrature of the average is not monotone in the number of nodes. The failure is aliasing of the oscillatory integrand against the node spacing: for a standard normal $Z$ and $k=4T\sigma$, both NumPy's and SciPy's 201-node rules give $E[\cos kZ]=0.80$ at $k=27.7$ where the exact value is $10^{-167}$, while 401 stable nodes give $10^{-14}$; at the signed echo's own $k=37$ the 201-node rule is accurate to $10^{-14}$ and the 401-node rule is wrong by $9\times10^{-3}$.

We therefore accept a quadrature only when its conditional probe maps agree with an independent reference built from Gaussian-weighted trapezoid rules of 481 and 961 points on $\pm6\sigma$ and 641 and 1281 points on $\pm8\sigma$. For the positive echo at $\sigma=0.05$ these references agree among themselves to $9\times10^{-16}$ when the spacing is halved and to $2.6\times10^{-10}$ when the cutoff is moved, and the accepted 101-node rule agrees with them to $1.2\times10^{-9}$.

For the signed echo the accepted rules (201 nodes at $\sigma=0.05$, 61 at $\sigma=0.02$) were compared directly with the same trapezoid references at both widths: at $\sigma=0.05$ the accepted 201-node rule agrees with the 1281-point reference on $\pm8\sigma$ to $1.1\times10^{-15}$ in the conditional maps and $3.6\times10^{-11}$ in $\partial_\theta p$, whereas the 61-node rule is off by $6.2\times10^{-4}$ at that width and is not used there; at $\sigma=0.02$ the accepted 61-node rule and the 201-node rule agree with the reference to $1.4\times10^{-15}$ and $7.7\times10^{-16}$ in the maps and to $3.3\times10^{-11}$ and $2.8\times10^{-12}$ in $\partial_\theta p$. Moving the reference cutoff from $\pm8\sigma$ to $\pm6\sigma$ changes it by $9.5\times10^{-10}$ ($\sigma=0.05$) and $2.8\times10^{-10}$ ($\sigma=0.02$), halving the spacing by less than $10^{-15}$. No node count is trusted without such a reference.

\section{Cycle, controller, estimator and the calibration design}
\label{app:cycle}

Each cycle consists of a decision interval of $0.05$, the pulse, the helper readout, an idle interval of $3$, and conditional resets of duration $0.4605$ (damping at rate 10 toward the sensor ground state at the true signal, removing 99\% of the excitation); each history starts from the maximally mixed sensor state and three resets. The helper is read along a preset equatorial axis at the end of the pulse and re-prepared in $\ket{+}$; neither operation is assigned a duration, so the charged time of a cycle has no separate readout term.

The controller evolves fifteen candidate sensor states, one for each pair of a plausible signal $\theta\in\{0.008,\dots,0.012\}$ and loading scale $s\in\{0.9,1,1.1\}$, conditioned on the observed outcomes, and acts on the worst candidate: before a pulse it forecasts the excitation the pulse would produce and resets if the forecast exceeds $10^{-3}$, repeating once and aborting the cycle if the forecast still exceeds it (the history is halted if, after the abort, the worst candidate's current excitation also exceeds $10^{-3}$); after the readout it resets if the conditional excitation exceeds $10^{-3}$, and halts the history if a second reset does not bring it down. No abort or halt occurred in any run reported.

The estimator is the posterior over a grid of 81 signal values (step $10^{-4}$) and nine (or three, for the echo runs) loading scales, propagated with the same conditional maps and reset actions. Both priors are uniform: the signal nodes carry equal weights, and the loading scale is marginalised with trapezoid weights on its nodes (end nodes at half weight); the conditional maps at the 81 signal values are cubic-spline interpolations of maps computed at 17 (echo: 9) signal nodes, with a minimum Choi eigenvalue of the interpolated maps checked at run time. The point estimate is the posterior mean of the signal; the 95\% credible interval is the central interval whose end points are the grid nodes at which the cumulative marginal posterior first reaches $0.025$ and $0.975$. Two readout conventions are used and should be kept apart. In every repeated-record simulation the helper readout angle is fixed at the no-offset optimum of the sequence, the equatorial angle that maximises the single-sequence Fisher information at the centre of the prior with the assignment error included: $2.313$ rad for the baseline pulse, $1.257$ rad for the signed echo, $0.833$ rad for the positive echo and $1.394$ rad for the plain comparator of the echo's duration; no run re-optimises the angle for the noise it is given. The single-sequence benchmarks of Sec.~\ref{sec:echo} and Fig.~\ref{fig:echo}(a), by contrast, report the information at the best angle for each noise setting; the re-optimisation moves the angle by about $0.01$ rad or less, and where both conventions were evaluated on the same sequence the information differs by less than one part in $10^4$, so the two conventions give the same single-sequence numbers at the quoted precision. Fisher information is accumulated along each history as $\sum_y(\partial_\theta p_y)^2/p_y$ from the conditional state and its derivative at the true signal and fixed loading scale; under drift this is conditional on the realised noise path. Rates are ratios of pooled totals over the 400 histories; uncertainties on rates and estimator statistics are bootstrap intervals over the histories, except the coverages under drift, which are quoted with Wilson intervals. Seeds are fixed and listed in the reproducibility package.

\begin{figure}
\centering\includegraphics[width=0.62\textwidth]{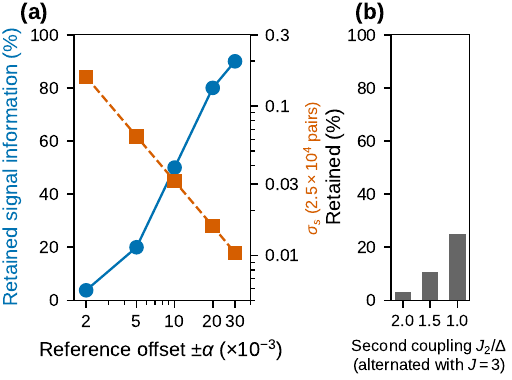}
\caption{Calibration design study in the ramp-integrated phase model with the squared visibility factor $0.728$ of the pulse. (a) Designs that add known reference offsets $\pm\alpha$ to the signal on alternate cycles: fraction of the signal information retained after adjusting for the unknown loading scale (circles, left axis) and loading-scale uncertainty after $2.5\times10^4$ pairs (squares, right axis). (b) Designs that alternate two coupling strengths, $J=3$ and a second value, without reference offsets: retained fraction of the signal information. These are local design predictions with preset readout quadratures and exactly known offsets.}
\label{fig:calibration}
\end{figure}

Because the record does not calibrate the coupling, we asked how it could be calibrated in place. The cleanest handle is a pair of known reference offsets $\pm\alpha$ added to the signal on alternate cycles: the branch phase $\Phi(\theta,s)$ depends on $\theta$ and $s$ differently, and two settings of $\theta$ separate them. To estimate the performance of the designs we use the phase of the actual pulse integrated over the ramps and the hold, $\Phi=\int\omega(\theta+\alpha,\,sJ(t))\,dt$, weighted by the squared visibility factor of the readout, $(1-2e)^2\e^{-2\gamma_MT}=0.728$ (a contrast amplitude of $0.853$). This analytic model reproduces the signal information of the full channel ($41\,973$ against $41\,950$) and its loading-scale information for the $\pm0.002$ pair to 8\%. Figure~\ref{fig:calibration} shows the trade-off. Offsets of $\pm0.02$ and $\pm0.03$ about $\theta_c$ retain 80\% and 90\% of the signal information after the nuisance adjustment and determine the loading scale to $0.016$ and $0.010$ after $2.5\times10^4$ pairs; smaller offsets retain little. Designs that alternate two coupling strengths instead do worse, retaining only $3.3$--$25\%$, because the ramps of the strong pulse add loading-scale sensitivity that is correlated with the signal. In all designs the phase stays within $1.03\,$rad of the preset readout quadrature over the prior, so no ambiguity arises. The reference offsets are taken as exactly known, the readout quadratures as separately preset, and the state carried from cycle to cycle is not modelled; a calibration run with uncertain references, and its coverage, is left for future work.

\section{Direct references, the QND instrument and the lifetime argument}
\label{app:direct}

The direct frontier of Fig.~\ref{fig:direct}(a) prepares the sensor by a five-unit pulse $u_x=a_x\sin^2(\pi t/5)$, $u_y=a_y\sin^2(\pi t/5)$ with bounded amplitudes, dresses it by a local $\sx$ field ramped at the same slew limit as the helper's coupling to $J_d\in\{0.5,1,1.5,2,3\}$, holds it for the remaining duration, unloads it, and reads it along the best axis on the Bloch sphere with assignment error $0.01$; the dissipator is rebuilt from the full timeline, the sensor is reset once per cycle, and the best $J_d$ is taken at each duration and noise setting.

The QND instrument of Fig.~\ref{fig:direct}(b) fixes the baseline-duration member of this family ($J_d=1.5$, $a_x=1.843$, $a_y=-0.487$) and runs it as a complete cycle (Table~\ref{tab:ops}). After the pulse an instantaneous ideal rotation, chosen once at $\theta_c$, brings the optimal binary-readout axis onto the energy basis of the design point, with projectors $\hat P_e$ and $\hat P_g$. Each read applies the instrument $\mathcal M_y(\rho)=(1-e)\hat P_y\rho\hat P_y+e\hat P_{1-y}\rho\hat P_{1-y}$ for the reported outcome $y$, whose trace is the reported-outcome probability and whose normalisation is the conditional state, followed by one time unit of evolution under the bath generator and the measurement dephasing $\mathcal D[\sqrt{\Gamma_m/2}\,(\hat P_e-\hat P_g)]$ with $\Gamma_m=10$.

A block of $m$ reads is accepted when all $m$ reports agree and a fresh block is taken otherwise, up to twelve blocks; every read and every rejected block is charged its duration. The probability of twelve consecutive rejections is below $10^{-20}$ per cycle for $m=2$, and the code would then proceed as for an accepted ``excited'' report. After acceptance the sensor is returned by an instantaneous swap of $\hat P_e$ and $\hat P_g$ if the accepted report is ``excited'' and left unchanged if it is ``ground''; this is a conditional basis swap, not the inverse of the analysis rotation. The Fisher information of each read is the expected information given the conditional state, summed over both reported outcomes, and the estimator is the posterior on the signal grid of Table~\ref{tab:settings}.

A one-read variant, which replaces verification by the helper's damping reset triggered on the excitation evaluated at the true signal, was also run; because that trigger uses the truth it is an oracle diagnostic, not an implementable controller, and it is not shown in Fig.~\ref{fig:direct}. Its rates are $324.8$, $295.9$ and $293.1$ at the three truths, below the two- and three-read instruments. The two-read rate at $\theta_c$ moves from $336$ to $314$ as the read duration goes from $0.1$ to $5$. At the duration of the signed echo ($T=185.1$, same dressing and preparation) the two- and three-read instruments collect $401.7$ and $399.9$ per unit time at the centre of the prior and $319.5$--$321.1$ at its corners, against the echo's $282.7$, $248.1$ and $256.5$.

\begin{figure}
\centering\includegraphics[width=0.62\textwidth]{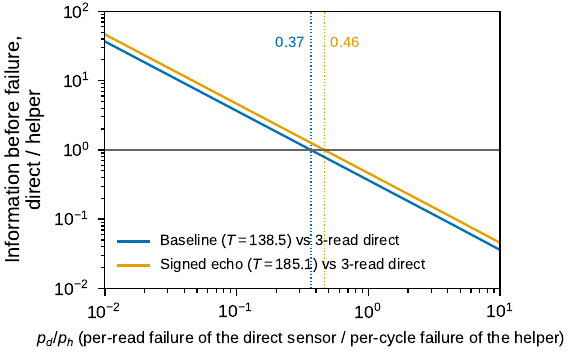}
\caption{Hypothetical lifetime comparison. Expected information collected before the first unmodelled failure of the sensor, direct relative to helper, as a function of the ratio of the direct sensor's failure probability per read, $p_d$, to the helper protocol's failure probability per cycle, $p_h$ ($p_h=10^{-3}$), for three-read verification at the baseline duration (blue) and at the duration of the echo (orange). Dotted lines mark the crossovers $p_d=0.37\,p_h$ and $0.46\,p_h$; the direct sensor wins to the left of its crossover and the helper to its right.}
\label{fig:lifetime}
\end{figure}

The lifetime argument of Sec.~\ref{sec:direct} supposes that each direct read has a probability $p_d$ of damaging the sensor in a way the model does not capture, and each helper cycle a probability $p_h$ of the same. The expected information collected before the first such event is $F(1-d)/d$ for a cycle of information $F$ and failure probability $d$; for $m$-read verification with assignment error $e$, independent read errors and a state frozen during the block, the per-cycle hazard is $d=1-a(1-p_d)^m/[1-(1-a)(1-p_d)^m]$ with $a=(1-e)^m+e^m$, obtained by summing the geometric series of rejected blocks. The two protocols break even at $p_d^\ast\simeq p_hF_d/(\bar N F_h)$, with $\bar N$ the mean number of reads per cycle: $p_d^\ast=0.37\,p_h$ for three-read verification at the baseline duration and $0.46\,p_h$ at the duration of the echo (Fig.~\ref{fig:lifetime}). The requirement is milder than it may look, because the verified direct sensor reads about three times per cycle. It is a hypothetical calculation without a device model.

\section{Drift sampling and the drift-averaged likelihood}
\label{app:drift}

For a stationary Ornstein--Uhlenbeck process of variance $\sigma^2$ and correlation time $\tau_c$, the end value $\beta'$ and the integral $I$ over a segment of duration $h$ given the start value $\beta$ are jointly Gaussian with $\beta'=a\beta+\eta$, $I=\tau_c(1-a)\beta+\zeta$, $a=\e^{-h/\tau_c}$, ${\rm Var}\,\eta=\sigma^2(1-a^2)$, ${\rm Cov}(\eta,\zeta)=\sigma^2\tau_c(1-a)^2$ and ${\rm Var}\,\zeta=\sigma^2\tau_c^2[2h/\tau_c-3+4a-a^2]$; the last expression is evaluated by its series for $h/\tau_c<0.05$ to avoid cancellation (worst relative error $3.2\times10^{-13}$ on the segment grid used). The process is advanced through the decision interval, any forecast reset, the first half of the sequence, the flip, the second half, the idle interval and any recovery reset.

The truth channel of a cycle is ${\rm Phase}(I_{\rm after})S_{\rm after}S_{\rm flip}(\beta_{\rm mid}){\rm Phase}(I_{\rm before})S_{\rm before}$ with $\beta_{\rm mid}$ the process value at the centre of the flip. The drift-averaged single-cycle channel is the expectation of this expression over the stationary joint law of $(I_{\rm before},\beta_{\rm mid},I_{\rm after})$, whose covariances follow from the exponential kernel; conditioned on $\beta_{\rm mid}$ the two integrals are Gaussian and their phase factors average analytically, $E[\e^{-\ii q^{\rm T}X}|b]=\exp(-\ii q^{\rm T}a\,b-q^{\rm T}Rq/2)$ with $a$ and $R$ the conditional mean coefficient and covariance, leaving a one-dimensional integral over $b$; a three-dimensional trapezoid product rule in the eigenbasis of the covariance agrees with this reduction to $3\times10^{-13}$ at $\tau_c=3\times10^3$ ($4\times10^{-11}$ at $3\times10^4$, where the covariance arithmetic limits the comparison).

The shrinkage rule follows from the small-phase limit of a readout probability $p=\tfrac12[1+VW\sin(\kappa(\theta-\theta_c))]$ fitted by a likelihood that assumes contrast $VW_{\rm a}$: the maximum sits at $\sin(\kappa(\hat\theta-\theta_c))=(W/W_{\rm a})\sin(\kappa(\theta-\theta_c))$. The adjacent-cycle covariance of the signed integrated offset of an ideal echo, with cycle separation $D$, is $-\sigma^2\tau_c^2\e^{-D/\tau_c}(\e^{T/2\tau_c}-1)^2(1-\e^{-T/2\tau_c})^2$, and the correlation at lag $m$ decays as $\e^{-(m-1)D/\tau_c}$.

For an ideal instantaneous echo the ensemble loss of phase information under the process is $1-\exp(-2\chi_E)$ with $\chi_E=4\sigma^2\tau_c^2[x-3+4\e^{-x/2}-\e^{-x}]$ and $x=T/\tau_c$, which gives $5.5\%$ and $42\%$ at $\tau_c=3\times10^4$ and $3\times10^3$. At the centre of the prior the nominal estimator's root-mean-square error under drift is $2.2$ and $2.0\times10^{-4}$ at the two correlation times against $2.35\times10^{-4}$ without drift, which is the shrinkage of Sec.~\ref{sec:echo} seen from the calibration point. For the plain protocol under the static shared offset the likelihood of a fixed record, at fixed $\theta$ and loading scale, is a trigonometric polynomial of degree at most 400 in the offset phase, periodic with period $\pi/T$, which 401 uniform nodes integrate exactly; the wrapped Gaussian prior is uniform over that period to $3\times10^{-12}$. The 401-node posterior's total-variation distance from the prior has median $1.3$--$1.4\times10^{-4}$ over the 100 histories at each truth and its relative entropy to the prior is at most $1.7\times10^{-5}$ nats; the nominal estimator's intervals there have width $7.8\times10^{-4}$. Echo filtering reduces the adjacent-cycle correlation of the residual phase noise to $-0.002$ and $-0.022$ at the two correlation times, against $0.99$ and $0.94$ for the raw offset. Under drift at the off-centre truth the maximum cycle-start excitation moved from $1.6\times10^{-4}$ without drift to $1.9$ and $2.1\times10^{-4}$, and the plain protocol of the echo's duration under the slower drift loses $37\%$ of its conditional information, resets once per 48 cycles, and its nominal estimator is overconfident (coverage $0.795$). The single-cycle midpoint diagnostic gives root-mean-square discrepancies of $1.96\times10^{-3}$ against a root-mean-square drift effect of $0.211$ on the readout probability.

The midpoint check draws 400 stationary paths at the faster drift and the off-centre truth, resolves the offset at every propagation step inside the flip, and compares the resulting sequence channel, conditional maps and readout probability with those of the midpoint model on the same path; the mean readout-probability discrepancy of $(2.40\pm0.97)\times10^{-4}$ and the root-mean-square discrepancy of $0.93\%$ of the root-mean-square per-cycle effect of the drift quoted in Sec.~\ref{sec:echo} are evaluated on the same paths and need not be uncorrelated (sample correlation $0.08$).

\end{appendices}

\bibliography{HAQS-LDL}

\end{document}